\documentclass[aps,prb,reprint,showpacs,citeautoscript]{revtex4-1}
\usepackage{hyperref, graphicx, graphics, verbatim, float, multirow, subfigure, amssymb,upgreek,array,lmodern}
\usepackage[utf8]{inputenc}
\usepackage[T1]{fontenc}

\begin{document}

\newcommand{\SixJSymbol}[6]{ \left\{
\begin{array}{c c c}
#1&#2&#3\\
#4&#5&#6\\
\end{array}
\right\}}

\title{Magnetic Dipole and Electric Quadrupole Transitions in the Trivalent Lanthanide Series: Calculated Emission Rates and Oscillator Strengths}
\author{Christopher M. Dodson}
\email[Email: ]{christopher\_dodson@brown.edu}
\author{Rashid Zia} 	
\email[Email: ]{rashid\_zia@brown.edu}
\affiliation{School of Engineering, Brown University, Providence, RI 02912 USA}
\date{\today}
\begin{abstract}
Given growing interest in optical-frequency magnetic dipole transitions, we use intermediate coupling calculations to identify strong magnetic dipole emission lines that are well suited for experimental study. The energy levels for all trivalent lanthanide ions in the $4f^{\text{n}}$ configuration are calculated using a detailed free ion Hamiltonian, including electrostatic and spin-orbit terms as well as two-body, three-body, spin-spin, spin-other-orbit, and electrostatically correlated spin-orbit interactions. These free ion energy levels and eigenstates are then used to calculate the oscillator strengths for all ground-state magnetic dipole absorption lines and the spontaneous emission rates for all magnetic dipole emission lines including transitions between excited states. A large number of strong magnetic dipole transitions are predicted throughout the visible and near-infrared spectrum, including many at longer wavelengths that would be ideal for experimental investigation of magnetic light-matter interactions with optical metamaterials and plasmonic antennas.\end{abstract}
\pacs{71.20.Eh,31.15.-p,32.50.+d}
\maketitle
\section{\label{sec:intro}Introduction}

The natural optical-frequency magnetic dipole (MD) transitions in trivalent lanthanide ions have attracted considerable attention in recent years for their ability to interact with the magnetic component of light.\cite{ThommenOptLett2006, NoginovaJAP2008, NoginovaOptExp2009,YangNanoToday2009, NiAPB2011, TaminiauNatComm2012, KaraveliOptLett2010, KaraveliPRL2011, FengOptLett2011, GrosjeanNanoLett2011, SheikholeslamiNanoLett2011} Although most light-matter interactions are mediated by electric fields through electric dipole (ED) transitions, the intra-$4f^{\text{n}}$ optical transitions of the lanthanide series are well-known to include strong MD contributions. \cite{FreedPhysRev1941, DrexhageProgressOptics1974, KunzPRB1980, RikkenPRL1995, ZampedriPRB2007, DuanPRB2011, WeberPhysRev1967, WeberPhysRev1968, CarnallJChemPhys1968} Spurred by recent advances in optical metamaterials and nanophotonics, researchers have proposed a variety of ways to leverage natural MD transitions, e.g. as the building blocks for homogeneous negative index materials \cite{ThommenOptLett2006} and as probes for the local magnetic field. \cite{NoginovaJAP2008, NoginovaOptExp2009,YangNanoToday2009, NiAPB2011,TaminiauNatComm2012} Experimental studies have also demonstrated how the competition between ED and MD processes can be used to achieve strong enhancement of MD emission \cite{KaraveliOptLett2010} and to broadly tune emission spectra.\cite{KaraveliPRL2011} Numerical investigations have shown how the enhanced magnetic field in and around metal and dielectric nanostructures can promote MD transitions, \cite{KlimovOptComm2002, KlimovPRA2000, FengOptLett2011, GrosjeanNanoLett2011, SheikholeslamiNanoLett2011, RollyPRB2012,SchmidtOptExp2012} illustrating how near-field enhancements can modify optical selection rules to promote higher order (ED forbidden) optical processes. \cite{ZuritaSanchezJOSAB2002a, ZuritaSanchezJOSAB2002b, KlimovLaserSpec2005, KlimovPRA2005, TojoPRL2004,TojoPRA2005,TojoPRAII2005, YangNanoToday2009, AndersenNaturePhysics2011, KernPRA2012, KlimovEPL2012,FilterPRB2012}

Recent studies have focused primarily on the visible $^5D_{0}\rightarrow{}^7F_{1}$ MD transition in trivalent Europium (Eu$^{3+}$)  and the near-infrared $^4I_{13/2}\rightarrow{}^4I_{15/2}$ MD transition in trivalent Erbium (Er$^{3+}$).\cite{ThommenOptLett2006, NoginovaJAP2008, NoginovaOptExp2009,NiAPB2011, TaminiauNatComm2012, KaraveliOptLett2010, KaraveliPRL2011} The emphasis on these transitions is not surprising, because they have a long history of scientific and technological importance. The $^5D_{0}\rightarrow{}^7F_{1}$ MD transition in Eu$^{3+}$ near $588$ nm was first characterized in 1941\cite{FreedPhysRev1941} and subsequently used by Drexhage,\cite{DrexhageProgressOptics1974} Kunz and Lukosz\cite{KunzPRB1980} in their authoritative studies of modified spontaneous emission. More recently, spontaneous emission from the Eu$^{3+}$ MD transition has served as a reference standard in studies of local field effects \cite{RikkenPRL1995, ZampedriPRB2007, DuanPRB2011} and ligand environments.\cite{WertsPCCP2002}  The Er$^{3+}$ $^4I_{13/2}\rightarrow{}^4I_{15/2}$ transition, emitting near $1.5$ $\upmu{}$m, is widely used for fiber amplifiers in optical telecommunication. The ED and MD contributions to this mixed transition were investigated as early as 1967 by Weber.\cite{WeberPhysRev1967,WeberPhysRev1968} More recently, Er$^{3+}$ has been used to demonstrate modifications in the local density of optical states \cite{SnoeksPRL1995} as well as stimulated emission along surface plasmon waveguides. \cite{AmbatiNanoLett2008} 

From an experimental perspective though, it would be helpful to identify additional MD transitions, especially in the near infrared range from $700-1000$ nm. As compared to the $588$ nm visible transition in Eu$^{3+}$, optical nanostructures are much easier to fabricate for longer wavelengths, and at longer wavelengths, plasmonic resonances also exhibit higher quality-factors due to lower Ohmic losses. In contrast to the 1.5 $\upmu$m line in Er$^{3+}$, transitions at wavelengths shorter than $1000$ nm can be readily observed with high efficiency using standard silicon photodetectors. 

Table 1 in the canonical paper by ~\citet{CarnallJChemPhys1968} has served as a definitive list of MD absorption lines for over 40 years, and since its publication, this table has been the basis for identifying possible MD transitions in various trivalent lanthanide ions. However, the use of this table to identify MD emission lines for experimental study suffers from two limitations. First and foremost, the table restricts itself to transitions involving ground state energy levels, and therefore, does not include potential MD transition lines that occur between excited states. Second, Ref.~\onlinecite{CarnallJChemPhys1968} limits the free ion Hamiltonian to only the electrostatic and spin-orbit interactions. More accurate values of the transition wavelengths, oscillator strengths, and spontaneous emission rates can be achieved by including higher order terms. 

In this paper, we explicitly calculate MD transitions over all possible excited energy levels in the trivalent lanthanide series. We also implement a more complex model for the free ion Hamiltonian, including not only the electrostatic and spin-orbit interactions but also two-body, three-body, spin-spin, spin-other-orbit, and electrostatically correlated spin-orbit interactions. This model is then used to identify all non-zero MD transitions, highlighting those lines that are most promising for experimental investigation. Using these results, we then analyze the effect of various host materials on the branching ratio of specific MD transitions. Additionally, calculations of electric quadrupole (EQ) transition rates and oscillator strengths have been carried out for completeness and to differentiate MDs from other higher order transitions.

\section{\label{secmodel}Method}

Calculations of MD transitions were made by first constructing a Hamiltonian for all $4f^{\text{n}}$ electron configurations. The free ion Hamiltonian used is of the form:\cite{CarnallJChemPhys1989}
\begin{eqnarray}
H_{FI}=&H_0&+\sum_{k=0,2,4,6}F^k f_k+\zeta_f A_{so}\nonumber{}\\*
&+& \alpha L(L+1)+\beta G(G_2)+\gamma G(R_7)\nonumber{}\\* 
&+& \sum_{i=2,3,4,6,7,8}T^i  t_i+\sum_{h=0,2,4}M^h m_h \nonumber{}\\*
&+&\sum_{f=2,4,6}P^f p_f .
\label{FreeIonH}
\end{eqnarray}
This Hamiltonian only considers valence electrons. The first term, $H_0$, denotes the central field Hamiltonian that shifts the absolute values of the energy levels but not their respective spacings. Given that the scope of this paper concerns transitions between levels, and their respective rates, calculations do not include $H_0$. For each subsequent term, the leading factor represents a radial fit parameter that is determined from experiment, while the trailing factor is an angular term that can be calculated explicitly from first principles. For instance, $F^k$ is the radial fit parameter for the electrostatic interaction, while $f_k$ is the calculated angular portion. The spin-orbit interaction is designated by $\zeta_f$ and $A_{so}$. $\alpha$, $\beta$, and $\gamma$ and their respective angular portions $L(L+1)$, $G(G_2)$, and $G(R_7)$ are the two-body interaction terms. Three-body interactions are denoted by $T^i$ and $t_i$. A combination of both the spin-spin and spin-other-orbit interactions are encompassed in the $M^h$ and $m_h$ terms. $P^f$ and $p_f$ denote the electrostatically correlated spin-orbit interaction. Note that this Hamiltonian does not include terms to account for crystal field effects. Although such terms are necessary in the calculations of intra-$4f^{\text{n}}$ ED transitions, they constitute only a small correction for MD and EQ transitions, which are directly allowed in intermediate coupling. Therefore, the values calculated here are representative quantities that can be used to predict and analyze MD transitions in any host material.

After constructing the angular terms using the methods outlined in Appendix A, we then used radial fit parameters tabulated in Ref.~\onlinecite{CarnallJChemPhys1989} to construct the full Hamiltonian matrix. This matrix was subsequently diagonalized to yield the free ion energy levels and the $\left|\psi{}[LS]J\right>$ eigenstates. $L$,$S$, and $J$ represent the total orbital, spin, and angular momenta, while we use $\psi$ to denote all other quantum numbers necessary to define each state. Note that we place $LS$ in brackets here to illustrate that they are no longer good quantum numbers; eigenstates in intermediate coupling are composed of a linear combination of different $LS$ terms with the same total angular momentum $J$. Following standard conventions, we label each level in Russell-Saunders ($^{2S+1}L_J$) notation according to their dominant $LS$ term(s). If no single $LS$ term has a fractional contribution greater than $50\%$, then we label the level according to the two largest $LS$ terms. Using the complete eigenstates, we perform subsequent calculations of oscillator strengths and transition rates between all levels. Thus, over the full trivalent lanthanide series ($4f^1-4f^{13}$), we consider a total of 192,177 possible transitions, see Table \ref{NumTerms}.
\squeezetable
\begin{table}[H]
\caption{Number of terms, levels, and total transitions for given $f^{\text{n}}$ configuration.}
\begin{tabular}{l| ccccccc}
\hline
\multirow{2}{*}{Configuration}&$f^1$&$f^2$&$f^3$&$f^4$&$f^5$&$f^6$&\multirow{2}{*}{$f^7$}\\
&($f^{13}$)&($f^{12}$)&($f^{11}$)&($f^{10}$)&($f^{9}$)&($f^{8}$) &\\
\hline
Number of&\multirow{2}{*}{1}&\multirow{2}{*}{7}&\multirow{2}{*}{17}&\multirow{2}{*}{47}&\multirow{2}{*}{73}&\multirow{2}{*}{119}&\multirow{2}{*}{119}\\
Terms ($LS$)&&&&&&&\\
\hline
Number of&\multirow{2}{*}{2}&\multirow{2}{*}{13}&\multirow{2}{*}{41} &\multirow{2}{*}{107}&\multirow{2}{*}{198}&\multirow{2}{*}{295}&\multirow{2}{*}{327}\\
Levels ($LSJ$)&&&&&&&\\
\hline
Number of&\multirow{2}{*}{1}&\multirow{2}{*}{78}&\multirow{2}{*}{820}&\multirow{2}{*}{5,671}&\multirow{2}{*}{19,503}&\multirow{2}{*}{43,365}&\multirow{2}{*}{53,301}\\
Transitions&&&&&&&\\
\hline
\end{tabular}
\label{NumTerms}
\end{table}
\section{\label{secresults}Results and Discussion}
\subsection{Magnetic Dipole Absorption Lines}

\begin{figure}
\includegraphics[scale=1]{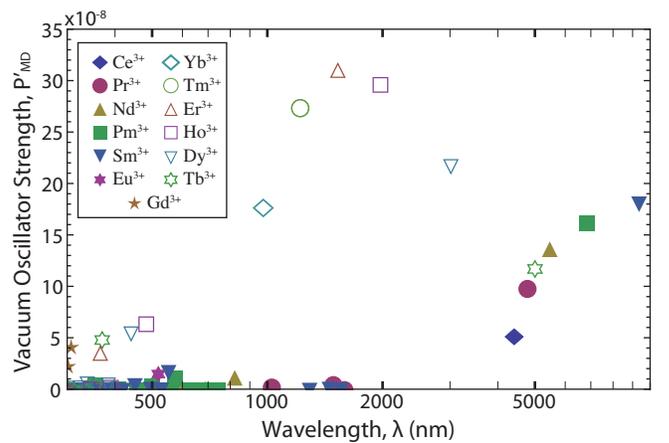}
\caption{(Color online) Plot of the magnetic dipole ground state absorption lines and corresponding MD oscillator strengths for all trivalent lanthanide ions between $300$ and $10000$ nm.}
\label{figStrongMDOscillators}
\end{figure}

We first calculate the oscillator strengths for all ground state MD absorption lines in the trivalent lanthanide series. (The formulas used for this calculation are provided in Appendix~\ref{MDTransitionsApp}.) Our results found $468$ non-zero MD absorption lines, including 84 transitions between $300$ nm and $10$ $\upmu{}$m;  the vacuum oscillator strengths, $P_{MD}^\prime{}$, of these transitions are plotted in Fig.~\ref{figStrongMDOscillators}. Table~\ref{TableMDOscillatorStrengths} shows a list of the most prominent ground state absorption lines, restricted to the energy bounds and minimum oscillator strengths used in Table 1 of ~\citet{CarnallJChemPhys1968}
\begin{table*}
\begin{ruledtabular}
\caption{Calculated MD vacuum oscillator strengths for trivalent lanthanides.\footnote{Only transitions with vacuum MD oscillator strength $P_{MD}^\prime{}>0.015$ are listed.}}
\begin{tabular}{c c c r r r r r |c c c r r r r r|}
& $SLJ$& $S^\prime L^\prime J^\prime$&\multicolumn{2}{c}{E(cm$^{-1}$)\footnotemark[2]}&$\lambda{}(nm)$&\multicolumn{2}{c}{$P_{MD}^\prime{}\times{}10^8$~\footnotemark[2]\footnotemark[3]}& &$SLJ$&$S^\prime L^\prime J^\prime$&\multicolumn{2}{c}{E(cm$^{-1}$)\footnotemark[2]}&$\lambda{}(nm)$&\multicolumn{2}{c}{$P_{MD}^\prime{}\times{}10^8$~\footnotemark[2]\footnotemark[3]}\footnotetext[2]{Italic values shown for comparison are taken from Table 1 of Ref~\onlinecite{CarnallJChemPhys1968}.}\footnotetext[3]{The MD oscillator strength, $P_{MD}$, inside a host material with refractive index $n_r$ would be: $P_{MD}=P_{MD}^\prime{}$ $n_r$}\\
\hline
Ce$^{3+}$ & $^2F_{5/2}$ & $^2F_{7/2}$ & 2266 &  & 4414 & 5.24 &  &  Gd$^{3+}$ & $^8S_{7/2}$ & $^6D_{9/2}$ & 39 524 & \emph{39779} & 253 & 0.04 & \emph{0.03}\\
\cline{1-8}
Pr$^{3+}$ & $^3H_4$ & $^3H_5$ & 2092 & \emph{2322} & 4781 & 9.86 & \emph{9.76} &  &  & $^6D_{7/2}$ & 40 647 & \emph{40712} & 246 & 0.55 & \emph{0.39}\\
 &  & $^3F_3$ & 6290 & \emph{6540} & 1590 & 0.02 & \emph{0.02} &  &  & $^6D_{5/2}$ & 40 928 & \emph{40977} & 244 & 0.29 & \emph{0.20}\\
\cline{9-16}
 &  & $^3F_4$ & 6720 & \emph{6973} & 1488 & 0.50 & \emph{0.49} & Tb$^{3+}$ & $^7F_6$ & $^7F_5$ & 1999 & \emph{2112} & 5003 & 11.90 & \emph{12.11}\\
 &  & $^1G_4$ & 9734 & \emph{9885} & 1027 & 0.27 & \emph{0.25} &  &  & $^5G_6$ & 27 004 & \emph{26425} & 370 & 5.01 & \emph{5.03}\\
\cline{1-8}
Nd$^{3+}$ & $^4I_{9/2}$ & $^4I_{11/2}$ & 1829 & \emph{2007} & 5468 & 13.75 & \emph{14.11} &  &  &  $^5G_5$ & 28 252 & \emph{27795} & 354 & 0.38 & \emph{0.36}\\
 &  & $^2H_{9/2}$ & 12 167 & \emph{12738} & 822 & 1.25 & \emph{1.12} &  &  & $^5L_6$ & 30 042 & \emph{29550} & 333 & 0.14 & \emph{0.14}\\
 &  & $^4F_{9/2}$ & 14 540 & \emph{14854} & 688 & 0.18 & \emph{0.20} &  &  & $^5H_7$ & 31 843 & \emph{31537} & 314 & 0.05 & \emph{0.06}\\
 &  & $^2G_{7/2}$ & 16 892 & \emph{17333} & 592 & 0.02 & \emph{0.02} &  &  & $^5H_6$ & 33 279 & \emph{33027} & 300 & 0.37 & \emph{0.46}\\
 &  & $^2G_{9/2}$ & 19 266 &  &  519 & 0.02 &  &  &  &  $^5H_5$ & 34 182 & \emph{33879} & 293 & 0.08 & \emph{0.03}\\
 &  & $^2I_{11/2}$ & 29 454 & \emph{28624} & 340 & 0.45 & \emph{0.05} &  &  & $^5F_5$ & 35 441 & \emph{34927} & 282 & 2.11 & \emph{1.87}\\
 &  & $^2H_{11/2}$ & 34 646 &  &  289 & 0.05 &  &  &  &  $^5G_6$ & 41 329 & \emph{41082} & 242 & 0.26  & \emph{0.23} \\
\cline{1-8}
Pm$^{3+}$ & $^5I_4$ & $^5I_5$ & 1462 & \emph{1577} & 6841 & 16.23 & \emph{16.36} &  &  & $(^5G$,$^5K)_5$ & 41 605 &  &  240 & 0.02  & \\
 &  & $^5F_{4}$ & 14 432 & \emph{14562} & 693 & 0.07 & \emph{0.08} &  &  & $^5K_6$ & 44 324 &  & 226 & 0.04 & \\
\cline{9-16}
 &  & $(^3H$,$^5G)_{4}$ & 17 376 & \emph{17327} & 575 & 1.23 & \emph{1.30} & Dy$^{3+}$ & $^6H_{15/2}$ & $^6H_{13/2}$ & 3316 & \emph{3506} & 3016 & 21.73 & \emph{22.68}\\
 &  & $^5G_{3}$ & 17 896 &  & 559 & 0.02 &  &  &  &  $^4I_{15/2}$ & 22 691 & \emph{22293} & 441 & 5.48 & \emph{5.95}\\
 &  & $^5G_{4}$ & 20 038 & \emph{20181} &  499 & 0.46 & \emph{0.26} &  &  & $^4K_{17/2}$ & 25 967 & \emph{26365} & 385 & 0.10 & \emph{0.09}\\
 &  & $^3G_{4}$ & 24 499 & \emph{23897} & 408 & 0.09 & \emph{0.11} &  &  & $^4I_{13/2}$ & 26 050 & \emph{25919} & 384 & 0.51 & \emph{0.41}\\
 &  & $^3G_{5}$ & 27 022 &  &  370 & 0.02 &  &  &  &  $^4M_{15/2}$ & 29 534 & \emph{29244} & 339 & 0.61 & \emph{0.69}\\
 &  & $^3I_{5}$ & 28 207 & \emph{27916} & 355 & 0.49 & \emph{0.23} &  &  & $^4M_{17/2}$ & 29 740 & \emph{30892} & 336 & 0.02 & \emph{0.03}\\
 &  & $^3H_{4}$ & 36 389 & \emph{35473} & 275 & 0.04 & \emph{0.04} &  &  & $(^4K$,$^4M)_{15/2}$ & 30 846 & \emph{31795} & 324 & 0.23 & \emph{0.12}\\
\cline{1-8}
Sm$^{3+}$ &  $^6H_{5/2}$ & $^6H_{7/2}$ & 1069 & \emph{1080} & 9355 & 18.12 & \emph{17.51} &  &  & $(^4K$,$^4L)_{13/2}$ & 33 321 & \emph{33776} & 300 & 0.20 & \emph{0.37}\\
 &  & $^6F_{3/2}$ & 6416 & \emph{6641} & 1559 & 0.03 & \emph{0.02} &  &  & $^4H_{13/2}$ & 33 924 & \emph{33471} & 295 & 1.41 & \emph{0.60}\\
 &  & $^6F_{5/2}$ & 6883 & \emph{7131} & 1453 & 0.11 & \emph{0.08} &  &  & $^4L_{15/2}$ & 36 261 &  &  276 & 0.02  & \\
 &  & $^4G_{5/2}$ & 18 116 & \emph{17924} & 552 & 1.73 & \emph{1.76} &  &  & $(^4L$,$^4K)_{13/2}$ & 36 666 &  & 273 & 0.02 & \\
 &  & $^4F_{3/2}$ & 18 918 & \emph{18832} & 529 & 0.03 & \emph{0.03} &  &  & $(^2K$,$^2L)_{15/2}$ & 38 434 & \emph{38811} & 260 & 0.15 & \emph{0.09}\\
\cline{9-16}
 &  & $^4G_{7/2}$ & 20 172 & \emph{20014} & 496 & 0.10 & \emph{0.05} & Ho$^{3+}$ & $^5I_8$ & $^5I_{7}$ & 5064 & \emph{5116} & 1975 & 29.72 & \emph{29.47}\\
 &  & $^4F_{5/2}$ & 22 177 & \emph{22098} & 451 & 0.45 & \emph{0.45} &  &  & $^3K_{8}$ & 20 715 & \emph{21308} & 483 & 6.46 & \emph{6.39}\\
 &  & $^4F_{7/2}$ & 24 889 & & 402 & 0.02 &  & & & $^3K_{7}$ & 25 636 & \emph{26117} & 390 & 0.28 & \emph{0.28}\\
 &  & $^4H_{7/2}$ & 28 715 & \emph{28396} & 348 & 0.04 & \emph{0.67} &  &  & $^3L_{9}$ & 28 873 & \emph{29020} & 346 & 0.14 & \emph{0.12}\\
 &  & $^4G_{5/2}$ & 30 079 & \emph{30232} & 332 & 0.04 & \emph{0.03} &  &  & $^3L_{8}$ & 33 577 & \emph{34206} & 298 & 0.21 & \emph{0.17}\\
 &  & $^4H_{7/2}$ & 42 572 & & 235 & 0.19 &  & & & $^3I_{7}$ & 37 258 & \emph{38470} & 268 & 0.24 & \emph{0.36}\\
\cline{9-16}
 &  & $^4G_{5/2}$ & 43 021 & \emph{42714} & 232 & 0.19 & \emph{0.02} & Er$^{3+}$ & $^4I_{15/2}$ & $^4I_{13/2}$ & 6534 & \emph{6610} & 1528 & 31.14 & \emph{30.82}\\
\cline{1-8}
Eu$^{3+}$ & $^7F_0$ & $^7F_{1}$ & 399 & \emph{350} & 25044 & 18.68 & \emph{17.73} &  &  & $^2K_{15/2}$ & 27 315 & \emph{27801} & 366 & 3.66 & \emph{3.69}\\
 &  & $^5D_{1}$ & 19 264 & \emph{19026} & 519 & 1.69 & \emph{1.62} &  &  & $^2K_{13/2}$ & 32 597 & \emph{33085} & 307 & 0.05 & \emph{0.11}\\
 &  & $^5F_{1}$ & 33 755 & \emph{33429} & 296 & 1.24 & \emph{2.16} &  &  & $^2L_{17/2}$ & 41 022 & \emph{41686} & 244 & 0.03 & \emph{0.03}\\
\cline{9-16}
 &  & $^3P_{1}$ & 38 891 &  & 257 & 0.05 &  &  Tm$^{3+}$ & $^3H_6$ & $^3H_{5}$ & 8205 & \emph{8390} & 1219 & 27.41 & \emph{27.25}\\
 &  & $^5D_{1}$ & 41 557 &  & 241 & 0.29 &  & & & $^2I_{6}$ & 34 212 & \emph{34886} & 292 & 1.42 & \emph{1.40}\\
\cline{1-8}\cline{9-16}
Gd$^{3+}$ & $^8S_{7/2}$ & $^6P_{7/2}$ & 32 557 & \emph{32224} & 307 & 4.28 & \emph{4.13} & Yb$^{3+}$ & $^2F_{7/2}$ & $^2F_{5/2}$ & 10 248 & \emph{10400} & 976 & 17.76 & \emph{17.76}\\
 &  & $^6P_{5/2}$ & 33 169 & \emph{32766} & 301 & 2.42 & \emph{2.33} &  &   &  &  &  &  \\
\end{tabular}
\label{TableMDOscillatorStrengths}
\end{ruledtabular}
\end{table*}

By comparison, we find 13 additional MD transitions that are not listed in Ref.~\onlinecite{CarnallJChemPhys1968}. While most of these new absorption lines are relatively weak, $P_{MD}^\prime{}\leq{}5\times10^{-10}$, several exhibit significant MD oscillator strengths, including the $^2F_{5/2}\rightarrow{}^2F_{7/2}$ ($4.14$ $\upmu$m) transition in Ce$^{3+}$, $^6H_{5/2}\rightarrow{}^4H_{7/2}$ ($235$ nm) transition in Sm$^{3+}$, and the $^7F_{0}\rightarrow{}^5D_{1}$ ($241$ nm) transition in Eu$^{3+}$ that have vacuum oscillator strengths of $5.24\times10^{-8}$, $1.9\times10^{-9}$ and $2.9\times10^{-9}$ respectively. As well as finding additional absorption lines, these calculations provide a more accurate prediction of transition wavelengths. For example, the $^4I_{15/2}\rightarrow{}^4I_{13/2}$ transition in Er$^{3+}$ is here calculated to occur at $1528$ nm, closer to the observed $1543$ nm center wavelength\cite{WeberPhysRev1967} than the $1513$ nm value reported in Ref.~\onlinecite{CarnallJChemPhys1968}. However, it is worth noting that the oscillator strengths are not significantly changed by the inclusion of higher order terms in the free ion Hamiltonian, as evidenced by the side-by-side comparison of $P_{MD}^\prime{}$ values in Table~\ref{TableMDOscillatorStrengths}. As further validation, our values also compare favorably with the Hartree-Fock code developed by R.D. Cowan and maintained at Los Alamos National Laboratory, \footnote{A web interface to the Cowan code is available at http://aphysics2.lanl.gov/tempweb/lanl/. Although this interface cannot be directly used to calculate MD oscillator strengths, the mixing coefficients for the intermediate coupling states can be used in combination with the expressions in Appendix~\ref{MDTransitionsApp} to calculate MD transition rates and strengths.} which predicts that the $^4I_{15/2}\rightarrow{}^4I_{13/2}$ transition in the $4f^{11}$ configuration of Er$^{3+}$ should occur at 1495.5 nm with an oscillator strength of 31.75$\times$10$^{-8}$, which is within 2\% of our calculated value of 31.14$\times$10$^{-8}$. For reference, a tabulated version of the all non-zero MD ground state absorption lines between $300$ and $10$ $\mu{}$m is provided in Table S1 of the Supplemental Material.\footnotemark[4]

\subsection{Magnetic Dipole Emission Lines} 
Beyond ground state absorption lines, there are MD transitions that occur solely between two excited states. Some of these excited transitions, such as the $^5D_0\rightarrow{}^7F_1$ transition in Eu$^{3+}$ and $^5D_4\rightarrow{}^7F_5$ transition in Tb$^{3+}$ have been identified experimentally.\cite{FreedPhysRev1941,TaminiauNatComm2012} However, there have been no exhaustive studies of MD emission in all trivalent lanthanide ions. Here, we use calculations to perform such a search. We proceed to tabulate all non-zero MD emission lines between $300$ and $1700$ nm. A total of 1927 non-zero MD emission lines were found throughout the lanthanide series. In Tables S2-S13 of the Supplemental Material we provide a complete list of all such transitions, grouping them by originating excited level to allow for a more convenient comparison in future experimental studies.\footnotemark[4] A more condensed table of strong transitions with vacuum emission rates, $A^\prime_{MD}$, greater than 5 $s^{-1}$ is shown in Table~\ref{TableMDEmissionLines}. 

\begin{table*}
\begin{ruledtabular}
\caption{Calculated MD vacuum spontaneous emission rates.\footnote{Only transitions between $300$ -- $1700$ nm with vacuum MD spontaneus emission rate $A_{MD}^\prime > 5$ $s^{-1}$ are listed.}}
\begin{tabular}{c c c r r | c c c r r|}
&$SLJ$&$S^\prime L^\prime J^\prime$&$\lambda$ (nm)&$A_{MD}^\prime{}$ $(s^{-1})$ \footnotemark[2]&&$SLJ$&$S^\prime L^\prime J^\prime$&$\lambda$ (nm)&$A_{MD}^\prime{}$ $(s^{-1})$ \footnotemark[2] \footnotetext[2]{The MD spontaneous emission rate, $A_{MD}$, inside a host material with refractive index $n_r$ would be: $A_{MD}=A_{MD}^\prime{}n_r^3$}\\
\hline
\text{Sm}$^{3+}$&$^4G_{11/2}$&$^6F_{11/2}$&477&7.14&Dy$^{3+}$&$(^4F,$$^4D)_{5/2}$&$^6F_{7/2}$&533&5.13 \\
& $^4D_{3/2}$&$^6F_{5/2}$&487&5.44&& $^6P_{3/2}$&$^6F_{5/2}$&555&8.89\\
& $^4D_{1/2}$&$^6F_{3/2}$&504&5.93&& $(^2K,$$^2L)_{15/2}$&$^4I_{15/2}$&635&9.75\\
\cline{1-5}
Eu$^{3+}$& $^5F_{2}$&$^7F_{1}$&304&5.49&& $(^4P,$$^6P)_{3/2}$&$^6F_{5/2}$&676&5.94\\
& $^5F_{4}$&$^7F_{5}$&336&5.62&& $^4F_{9/2}$&$^6F_{11/2}$&734&11.72\\
& $^5F_{5}$&$^7F_{6}$&339&5.44&& $^4G_{9/2}$&$^4G_{11/2}$&896&6.64\\
& $^5D_{4}$&$^7F_{5}$&417&5.47&& $(^2K$,$^2L)_{15/2}$&$^4M_{15/2}$&1124&8.33\\
& $^5G_{4}$&$^7F_{4}$&418&6.31&&$(^4P$,$^4D)_{3/2}$&$^6P_{3/2}$&1170&5.19\\
& $^5G_{5}$&$^7F_{5}$&436&8.30&& $^4G_{9/2}$&$^4G_{11/2}$&1550&6.21\\
\cline{6-10}
& $^5G_{6}$&$^7F_{6}$&455&10.51&Ho$^{3+}$&$(^3H$,$^3G)_{5}$&$^5I_{6}$&361&12.20\\
& $^5D_{3}$&$^7F_{4}$&460&9.02&& $(^1G$,$^3H)_{4}$&$^5I_{5}$&411&9.86\\
& $^5D_{2}$&$^7F_{3}$&505&11.58&& $^3F_{4}$&$^5F_{5}$&422&6.73\\
& $^5D_{1}$&$^7F_{2}$&550&12.29&& $^3H_{6}$&$^5I_{7}$&449&24.71\\
& $^3P_{1}$&$^5D_{2}$&583&16.01&&$(^1G$,$^3H)_4$&$^5I_4$&449&5.05\\
& $^5D_{0}$&$^7F_{1}$&584&14.37&& $^3K_{8}$&$^5I_{8}$&483&18.48\\
& $^3P_{0}$&$^5D_{1}$&700&24.63&&$^3F_4$&$^5F_4$&486&5.71\\
& $(^3I$,$^3H)_6$&$^5G_{6}$&776&5.51&& $^3K_{7}$&$^5I_{7}$&486&8.68\\
\cline{1-5}
Gd$^{3+}$& $^6P_{5/2}$&$^8S_{7/2}$&301&23.64 && $^3P_{2}$&$^5S_{2}$&511&6.61\\
& $^6P_{7/2}$&$^8S_{7/2}$&307&30.24&& $^3F_{4}$&$^5F_{3}$&538&6.47\\
\cline{1-5}
Tb$^{3+}$& $^5F_{3}$&$^7F_{3}$&306&6.65&& $(^5G$,$^3H)_{5}$&$^5I_{6}$&543&8.35\\
& $^5F_{2}$&$^7F_{2}$&308&10.07&& $(^3F$,$^3G)_{4}$&$^5I_{4}$&618&7.72\\
& $^5F_{1}$&$^7F_{1}$&310&14.40&& $^3F_{4}$&$(^5G$,$^3G)_{5}$&653&16.60\\
& $^5F_{1}$&$^7F_{0}$&312&8.81&& $^3P_{1}$&$^5S_{2}$&661&6.19\\
& $^5G_{6}$&$^7F_{6}$&370&24.35&& $^3D_{3}$&$^5F_{4}$&672&12.32\\
& $^5D_{0}$&$^7F_{1}$&378&29.24&& $^3L_{8}$&$^3K_{8}$&777&11.42\\
& $^5G_{5}$&$^7F_{5}$&381&14.54&& $^3P_{1}$&$^5F_{2}$&800&5.00\\
& $^5D_{1}$&$^7F_{2}$&381&20.20&& $^3L_{7}$&$^3K_{7}$&811&5.60\\
& $^5D_{1}$&$^7F_{0}$&392&8.21&& $(^3H$,$^3G)_{5}$&$(^5G$,$^3H)_{5}$&1078&6.22\\
& $^5G_{4}$&$^7F_{4}$&393&9.68&& $(^3H$,$^3G)_{5}$&$^3H_{6}$&1126&12.12\\
& $^5G_{3}$&$^7F_{3}$&399&5.57&& $(^5F$,$^5G)_{2}$&$^5F_{3}$&1270&6.40\\
& $^5D_{2}$&$^7F_{3}$&409&17.88 && $(^5D$,$^5G)_4$&$(^5G$,$^3H)_5$&$1438$&$6.48$\\
 \cline{6-10}
& $^5D_{3}$&$^7F_{4}$&427&15.49&Er$^{3+}$&$^2K{}_{15/2}$& $^4I{}_{15/2}$ &$366$&$18.20$ \\
& $^5D_{2}$&$^7F_{1}$&430&7.11&$$&$^2K{}_{13/2}$& $^4I{}_{13/2}$ &$384$&$5.25$ \\
& $^5D_{4}$&$^7F_{5}$&530&14.32&$$&$(^2H$,$^2G)_{9/2}$& $^4I{}_{11/2}$ &$392$&$5.34$ \\
& $(^5D$,$^3P)_2$&$^5D_{3}$&766&17.49&$$&$^4G{}_{11/2}$& $^4I{}_{13/2}$ &$529$&$12.36$ \\
\cline{1-5}
Dy$^{3+}$& $(^4G$,$^4P)_{5/2}$&$^6H_{7/2}$&334&5.71&$$&$^2D{}_{5/2}$& $^4F{}_{7/2}$ &$583$&$5.05$ \\
& $^4G_{7/2}$&$^6H_{9/2}$&347&8.28&$$ &$^2P{}_{1/2}$& $^4S{}_{3/2}$&$668$&$11.31$ \\
& $(^4G$,$^4P)_{5/2}$&$^6H_{5/2}$&348&5.58&$$ &$^2D{}_{5/2}$& $^4F{}_{7/2}$&$686$&$20.05$ \\
& $^4H_{7/2}$&$^6H_{5/2}$&360&12.78&$$&$(^2G$,$^4F)_{9/2}$& $^4I{}_{11/2}$ &$733$&$8.48$\\
& $^4G_{11/2}$&$^6F_{11/2}$&361&15.99&& $(^2P$,$^2D)_{3/2}$&$^4S_{3/2}$&764&5.55\\
&$(^4H$,$^4G)_{9/2}$&$^6F{}_{9/2}$&$362$&$6.15$&& $(^2H$,$^4G)_{11/2}$&$^4I_{13/2}$&832&14.86\\
& $^4G{}_{7/2}$&$^6H{}_{7/2}$&$366$&$6.84$&& $(^2H$,$^2G)_{9/2}$&$(^2G$,$^4F)_{9/2}$&843&11.21\\
& $(^4H$,$^4G)_{11/2}$&$^6F{}_{11/2}$&$375$&$5.74$&& $(^2H$,$^2G)_{9/2}$&$^4G_{11/2}$&978&11.87\\
& $(^4P{}$,$^4D{})_{3/2}$&$^6F{}_{5/2}$ &$376$&$6.62$&& $(^2P$,$^2D)_{3/2}$&$^4F_{5/2}$&1081&8.19\\
& $^4G{}_{9/2}$&$^6F{}_{9/2}$&$386$&$8.45$&& $(^2G$,$^4F)_{9/2}$&$^4F_{9/2}$&1101&10.35\\
& $^4D{}_{7/2}$&$^6F{}_{9/2}$ &$400$&$9.44$&& $(^2P$,$^2D)_{3/2}$&$^4F_{3/2}$&1111&8.56\\
& $^4G{}_{9/2}$& $^6H{}_{11/2}$ &$410$&$11.35$&& $^4G_{9/2}$&$(^2H$,$^4G)_{11/2}$&1276&12.21\\
& $^4P{}_{1/2}$&$^6F{}_{3/2}$&$412$&$9.28$&& $^4I_{13/2}$&$^4I_{15/2}$&1528&10.17\\
& $(^4G{}$,$^4F{})_{7/2}$&$^6H{}_{9/2}$ &$415$&$8.77$&& $^4G_{7/2}$&$^4G_{9/2}$&1533&6.43\\
\cline{6-10}
& $^4P_{1/2}$&$^6F_{1/2}$&421&6.67&Tm$^{3+}$&$(^3P$,$^1D)_2$&$^3F_3$&430&22.93\\
& $(^4F$,$^4G)_{5/2}$&$^6F_{7/2}$&428&6.72&& $(^3P$,$^1D)_2$&$^3F_3$&765&13.97\\
& $^4I_{11/2}$&$^6H_{11/2}$&436&5.71& &$^1G_4$&$^3H_5$&784&22.64\\
& $^4I_{13/2}$&$^6H_{13/2}$&440&9.99&& $(^3P$,$^1D)_2$&$^3F_2$&808&9.29\\
& $^4I_{15/2}$&$^6H_{15/2}$&441&18.83&&$(^3P$,$^1D)_2$&$(^3P$,$^1D)_2$&983&12.96\\
& $(^4D$,$^4G)_{5/2}$&$^6H_{5/2}$&458&5.09&&$^3F_3$&$^3F_4$&1155&10.88\\
& $(^4D$,$^4G)_{5/2}$&$^6F_{7/2}$&471&8.11&& $^1G_4$&$^3H_4$&1167&5.60\\
& $^4G_{11/2}$&$^6H_{13/2}$&493&19.49&& $^3H_5$&$^3H_6$&1219&14.55\\
\cline{6-10}
& $^4F_{3/2}$&$^6F_{1/2}$&495&7.27&Yb$^{3+}$&$^2F_{5/2}$&$^2F_{7/2}$&976&16.59\\
& $(^4D$,$^4P)_{1/2}$&$^6F_{3/2}$&530&10.38&&&&&\\
\end{tabular}
\label{TableMDEmissionLines}
\end{ruledtabular}
\end{table*}

\begin{figure}
\includegraphics[scale=.97]{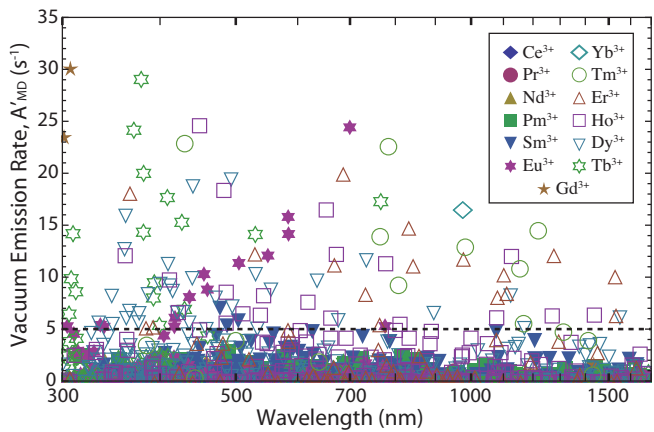}
\caption{(Color online) Magnetic dipole emission lines and corresponding vacuum emission rates for all trivalent lanthanide ions between $300$ and $1700$ nm. Strong emission lines with vacuum rates greater than 5 $s^{-1}$ located above the dashed line are listed in Table III.}
\label{figStrongVisTransitions}
\end{figure}

As shown in Figure~\ref{figStrongVisTransitions}, there are many strong MD transitions thoughout the ultraviolet, visible, and near infrared spectra. In addition to transitions which have been previously identified through ground state calculations or experimental characterization, there are many more MD emission lines which could be of practical interest. 

In the ultraviolet spectrum, MD transitions in Er$^{3+}$, Gd$^{3+}$, and Tb$^{3+}$ are particularly strong. The $^6P_{5/2}\rightarrow{}^8S_{7/2}$ ($301$ nm) and $^6P_{7/2}\rightarrow{}^8S_{7/2}$ ($307$ nm) transitions in Gd$^{3+}$ have vacuum emission rates of 23.64 and 30.24 $s^{-1}$, respectively. Similarly, the $^2K_{15/2}\rightarrow{}^4I_{15/2}$ ($366$ nm) transition in Er$^{3+}$ has a vacuum emission rate of 18.20 $s^{-1}$. Note that these transitions to the  $^4I_{15/2}$ ground state in Er$^{3+}$ and the $^8S_{7/2}$ ground state in Gd$^{3+}$ could have been inferred from the absorption lines discussed in the previous section. However, the strong UV transitions in Tb$^{3+}$ occur between excited states, such as the $^5D_{0}\rightarrow{}^7F_{1}$ ($378$ nm) and $^5D_{1}\rightarrow{}^7F_{2}$ ($381$ nm) which have vacuum emission rates of 29.24 and 20.20 $s^{-1}$, respectively. These $^5D_{J}\rightarrow{}^7F_{J+1}$ Tb$^{3+}$ transitions are the higher level analogues to the experimentally characterized  $^5D_{4}\rightarrow{}^7F_{5}$ ($530$ nm) excited state transition.

Throughout the visible spectrum, there are strong MD transitions in Eu$^{3+}$, Ho$^{3+}$, and Tb$^{3+}$. Similar to the UV transitions in Tb$^{3+}$, many of the visible MD transitions in Eu$^{3+}$ and Tb$^{3+}$ are higher level analogues to the previously known $^5D_{J}\rightarrow{}^7F_{J+1}$ transitions. In Eu$^{3+}$, the well-known $^5D_{0}\rightarrow{}^7F_{1}$ ($584$ nm) transition has a calculated vacuum emission rate of 14.37 $s^{-1}$.  In addition to this yellow emission line, there are also higher energy blue and green MD transitions in Eu$^{3+}$, including the $^5D_{1}\rightarrow{}^7F_{2}$ ($550$ nm), $^5D_{2}\rightarrow{}^7F_{3}$ ($505$ nm), and
$^5D_{3}\rightarrow{}^7F_{4}$ ($460$ nm) that have vacuum emission rates near 10 $s^{-1}$ each. Likewise, in addition to the green $^5D_{4}\rightarrow{}^7F_{5}$ ($530$ nm) line and higher ultraviolet transitions, Tb$^{3+}$ also has several blue-violet MD transitions, such as $^5D_{2}\rightarrow{}^7F_{3}$ ($409$ nm) and $^5D_{3}\rightarrow{}^7F_{4}$ ($420$ nm) which have vacuum emission rates greater than 15 $s^{-1}$. Trivalent Holmium (Ho$^{3+}$) also exhibits several strong blue MD transitions. Two prominent  Ho$^{3+}$ transitions are the $^3K_{8}\rightarrow{}^5I_{8}$ ($483$ nm) ground state transition and the  $^3H_{6}\rightarrow{}^5I_{7}$
($449$ nm) excited state transition, which have vacuum emission rates of 18.48 and 24.71 $s^{-1}$, respectively.

Most interestingly from an experimental perspective, there are also many strong MD transitions in the near-infrared spectrum. At these longer wavelengths, the design and fabrication of metamaterials,\cite{ZhangPRL2005,EnkrichPRL2005, ValentineNature2008,SersicPRL2009,AmelingNanoLett2010,XiaoNature2010,SoukoulisNatPhoton2011} resonant optical antennas,\cite{NovotnyPRL2007,CurtoScience2010,DregelyNatComm2011,BarnardNatNano2011} photonic crystals,\cite{BurresiPRL2010,VignoliniPRL2010} and plasmonic waveguides\cite{CharbonneauOptLett2000,WeeberPhysRevB2003,WeeberAPL2005,DitlbacherPRL2005,BozhevolnyiNature2006,ZiaPRB2006,BuckleyOptExp2007} are more established. Although some transitions in this regime originate from excited states that would require deep UV excitation, there are a number of transitions in Dy$^{3+}$, Er$^{3+}$, Tm$^{3+}$, and Yb$^{3+}$ that can be pumped at visible or near-IR wavelengths and are thus strong candidates for experimental use. These include several ground state transitions that could be identified from the absorption line calculations in the previous section, including the $^4I_{15/2}\rightarrow{}^4I_{13/2}$ ($1528$ nm) transition in Er$^{3+}$, the $^3H_5\rightarrow{}^3H_6$ ($1219$ nm) transition in Tm$^{3+}$, and the $^2F_{5/2}\rightarrow{}^2F_{7/2}$ ($976$ nm) transition in Yb$^{3+}$. Here, we calculate the MD vacuum emission rates of these transitions to be 10.17, 14.55, and 16.59 $s^{-1}$, respectively. Our calculations also reveal several promising excited state MD transitions. These include the $^4F_{9/2}\rightarrow{}^6F_{11/2}$ ($734$ nm) transition in Dy$^{3+}$, the $^1G_4\rightarrow{}^3H_5$ ($784$ nm) transition in Tm$^{3+}$, and the $(^2H,^4G)_{11/2}\rightarrow{}^4I_{13/2}$ ($832$ nm) transition in Er$^{3+}$ that have vacuum emission rates of 11.72, 22.64, and 14.86  $s^{-1}$, respectively. 

\begin {table*}
\centering
\begin{minipage}{5.15in}
\caption {Calculated MD Branching Ratios for Yb$^{3+}$ Transition in Different Host Materials}
\begin{ruledtabular}
\begin {tabular} {l | c | c | c | c}
Host & Measured Lifetime \footnotemark[1] \footnotetext[1]{From Table III in Ref. \onlinecite{DeLoachIEEEJQE1993}} & Refractive Index \footnotemark[2] \footnotetext[2]{From Table II in Ref. \onlinecite{DeLoachIEEEJQE1993}} & MD Emission Rate &  MD Branching Ratio \\
 & $\tau$ (ms)  & $n_r$ & $A_{MD}$ (s$^{-1}$) &   $\beta_{MD}$\\
\hline
LiYF$_4$ & 2.16 & 1.455 & 51.10 & 11.0\%\\
LaF$_3$ &  2.22 & 1.597 & 67.57 & 15.0\%\\
SrF$_2$ &  9.72 & 1.438 & 49.33 & 48.0\%\\
BaF$_2$ &  8.2 & 1.473 & 53.02 & 43.5\%\\
KCaF$_3$&  2.7 & 1.378 & 43.41 & 11.7\%\\
KY$_3$F$_10$ &  2.08 & 1.5 & 55.99 & 11.6\%\\
Rb$_2$NaYF$_6$ &  10.84 & 1.403 & 45.82 & 49.7\%\\
BaY$_2$F$_8$ &   2.04 & 1.521 & 58.38 & 11.9\%\\
Y$_2$SiO$_5$ &  1.04 & 1.79 & 95.15 & 9.9\%\\
Y$_3$Al$_5$O$_{12}$ &  1.08 & 1.82 & 100.0 &  10.8\%\\
YAIO$_3$ &  0.72 & 1.956 & 124.2 & 8.9\%\\
Ca$_5$(PO$_4$)$_3$F &  1.08 & 1.63 & 71.85 & 7.8\%\\
LuPO$_4$ &  0.83 & 1.83 (est.) & 101.7 & 8.4\%\\
LiYO$_2$ &  1.13 & 1.82 (est.) & 100.0 & 11.3\%\\
ScBO$_3$ &  4.8 & 1.84 & 103.3 & 49.6\%\\
\end {tabular}
\label {YbAnalysis}
\end{ruledtabular}
\end{minipage}
\end {table*}

Of the seven strong near-infrared lines identified above, the four transitions between 700 and 1000 nm are the most promising candidates for immediate experimental study. Unlike longer wavelength transitions such as the 1.5 $\upmu$m transition in Er$^{3+}$, these MD transitions occur in a spectral region where they are still readily observed by silicon photodetectors. (For example, back-illuminated CCD cameras such as the Pixis 1024B from Princeton Instruments exhibit greater than 50\% quantum efficiency up to 900nm.) Nevertheless, these transitions also occur at sufficiently long wavelengths that resonant plasmonic and nanophotonic structures can be readily fabricated.

For experimental studies, it will also be important to select appropriate host materials to maximize MD emission. In particular, to enhance the MD contribution to mixed transitions, it will be helpful for lanthanide ions to be substitutionally doped into centrosymmetric sites where ED transitions are strictly forbidden. Table \ref{YbAnalysis} shows the calculated MD branching ratios for the Yb$^{3+}$ $^2F_{5/2}\rightarrow{}^2F_{7/2}$ ($976$ nm) transition in different host materials. These calculations were performed by comparing the total decay rate ($\Gamma_{total}= 1/{\tau}$), as inferred from experimental lifetime data in the literature,\cite{DeLoachIEEEJQE1993} with the MD spontaneous emission rates  ($A_{MD}=A_{MD}^\prime{}$ $n_r^3$)\cite{RikkenPRL1995, ZampedriPRB2007, DuanPRB2011} predicted from the vacuum rates in Table \ref{TableMDEmissionLines}.\footnote{Note that crystal field effects in each host will also introduce a perturbation to the vacuum emission rates. However, for MD transitions, this effect is generally much smaller than the optical density of states correction due to the host's refractive index. As a quantitative example, we have explicitly calculated the MD emission rates for the Yb$^{3+}$ $^2F_{5/2}\rightarrow{}^2F_{7/2}$ transitions in LaF$_3$ by adding crystal field corrections from Ref.~\onlinecite{CarnallJChemPhys1989} to our detailed free ion Hamiltonian. Although there is minor variation in the vacuum emission rate for different states within the $^2F_{5/2}$ level (i.e. different excited states with M$_J$=$\pm$5/2, $\pm$3/2, $\pm$1/2), the maximum deviation from the free ion vacuum emission rate is less than 20\%. In contrast, the refractive index effect of the LaF$_3$ host, which scales as n$_r^3$, enhances the MD emission rate by over 300\%.} The MD branching ratio is thus defined as: $\beta_{MD} = A_{MD}/{\Gamma_{total}}$. Note that the MD branching ratio for this Yb$^{3+}$ transition varies significantly in different host materials.  In centrosymmetric hosts such as SrF$_2$, Rb$_2$NaYF$_6$, and ScBO$_3$, it is possible to have $\sim$50\% of all decay processes result in MD emission. In more common materials, such as yttrium aluminum garnet (YAG, Y$_3$Al$_5$O$_{12}$), MD emission still accounts for $\sim$10\% of all decay processes. 

The relatively simple two-level energy structure of Yb$^{3+}$ means that MD emission can naturally account for a significant contribution to the overall decay. Other, more complex energy level structures, such as in Dy$^{3+}$ and Tm$^{3+}$, mean that there are more decay paths from any particular excited state. These transitions are thus interesting candidates for enhancing MD emission. For instance, the lifetime of the $^4F_{9/2}$ excited level in Dy$^{3+}$ ranges from 300 $\upmu$s in LiNbO$_3$ \cite{DominiakDzikJMolStruc2004} to 2.36 ms in Y$_3$Sc$_2$Ga$_3$O$_{12}$ (YSGG) \cite{SardarJlumi2004} leading to respective branching ratios of 0.35\% and 2.77\% for the associated $^4F_{9/2}\rightarrow{}^6F_{11/2}$ MD transition. Similar branching ratios were found by analyzing the $^1G_4\rightarrow{}^3H_5$ transition in Tm$^{3+}$.\cite{GueryJofLumi1988,PisarskiJPhysCondensMat2004,SokolskaJPhysChemSolids2000}
\subsection{Electric Quadrupole Calculations}

In the multipolar expansion of light-matter interactions, MD terms are generally included in the same order as EQ terms, because they both scale with spatial derivatives of the electric field. Thus, a common question is to what extent EQ transitions compete with MD transitions. For completeness, we have calculated the oscillators strengths for all EQ ground state absorption lines and the spontaneous emission rates for all EQ emission lines. The EQ oscillator strengths and transition rates were found to be significantly smaller than those for MD transitions. 
\begin{figure}[!]
\includegraphics[scale=1]{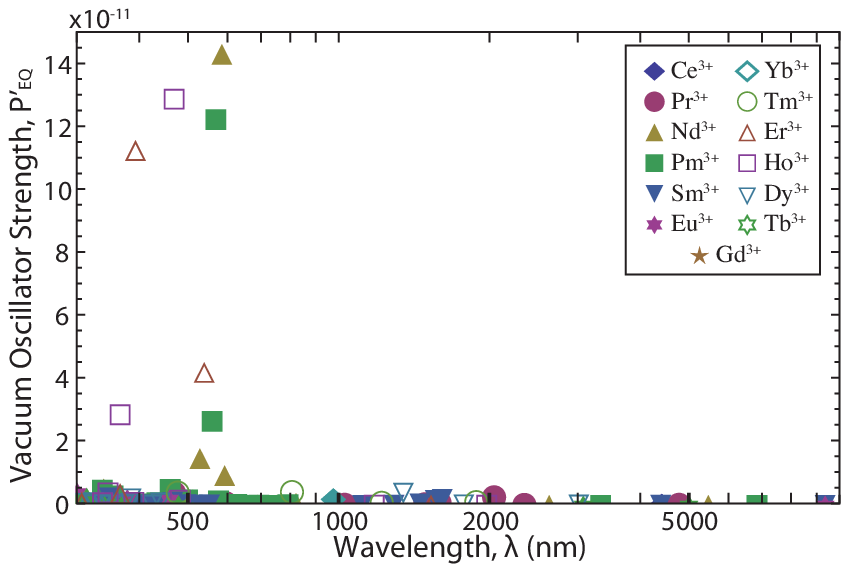}
\caption{(Color online) Plot of all electric quadruple ground state absorption lines and corresponding EQ oscillator strengths for all trivalent lanthanide ions between $300$ and $10000$ nm.}
\label{figEQOscillators}
\end{figure}

\begin{figure}[!]
\includegraphics[scale=.97]{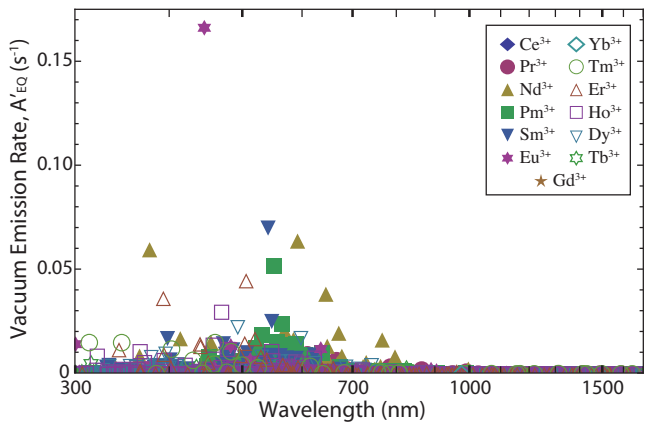}
\caption{(Color online) Plot of all electric quadrupole emission lines and corresponding EQ vacuum emission rates for all trivalent lanthanide ions between $300$ and $1700$ nm.}
\label{figEQEmissionLines}
\end{figure}

The strongest EQ transition was the $(^5D,^5P)_2\rightarrow{}^5D_0$ transition in Eu$^{3+}$ with a vacuum emission rate of $0.17$ $s^{-1}$. While the emission rate for EQ transitions scales with $n_r^5$, this rate is approximately 30 times weaker than the weakest MD transition presented in Table~\ref{TableMDEmissionLines}. Most transitions mediated by EQ interactions have an emission rate on the order of $0.01$ s$^{-1}$ and would thus require significant enhancement to even be observed. Figures \ref{figEQOscillators} and  \ref{figEQEmissionLines} show the vacuum oscillator strengths and emission rates, respectively, for EQ absorption lines and EQ emission lines. A complete tabulation of all 236 EQ absorption lines (Table S14) and all 3079 EQ emission lines (Tables S15-S25) between $300$ and $1700$ nm is provided in the Supplemental Material.\footnotemark[4] These calculations confirm that EQ transitions in trivalent lanthanide ions are negligible in comparison to the MD transitions calculated above.

\section{\label{secconclusion}Conclusion}

Using a detailed free ion Hamilitonian, we have calculated all non-zero MD ground state absorption lines and corresponding oscillator strengths throughout the full trivalent lanthanide series. These values are well documented in the literature, and we observed good agreement between our results and those found in Ref.~\onlinecite{CarnallJChemPhys1968}. Using this detailed Hamiltonian, we then calculated all non-zero MD and EQ emission lines and their respective emission rates for all trivalent lanthanide ions. Although the EQ emission rates were found to be negligible, our calculations revealed vastly more MD emission lines than previously identified by ground state calculations or experimental investigation.  

In the specific spectral range from 300 -- 1700 nm, we identified 1927 MD transitions, including 117 lines with vacuum spontaneous emission rates $A^\prime{}_{MD}>$ 5 s$^{-1}$.  Of these transitions, four were identified as the most promising for experimental exploration: $^4F_{9/2}\rightarrow{}^6F_{11/2}$ ($734$ nm) in Dy$^{3+}$, $^1G_4\rightarrow{}^3H_5$ ($784$ nm) in Tm$^{3+}$, $(^2H,^4G)_{11/2}\rightarrow{}^4I_{13/2}$ ($832$ nm) in Er$^{3+}$, and $^2F_{5/2}\rightarrow{}^2F_{7/2}$ ($976$ nm) in Yb$^{3+}$. These near-IR transitions occur at wavelengths for which resonant devices are easily fabricated, yet still emit within the detection range of silicon photodetectors. 

We subsequently demonstrated how free ion calculations can be used to analyze and predict MD emission within a range of host materials. We compared the calculated emission rates with experimental lifetime data from the literature to approximate MD branching ratios, and for the specific case of the $^2F_{5/2}$ excited level in Yb$^{3+}$, showed how MD emission can account for up to $\sim$50\% of all decay processes. These calculations highlighted the importance of selecting appropriate hosts, especially those with high centrosymmetry and refractive indices, to maximize MD contributions.

These results and the associated tables in the Supplemental Material\footnotemark[4] can thus be used to guide the study of magnetic light-matter interactions in trivalent lanthanide ions. Beyond the well-known MD emission lines in Eu$^{3+}$ and Er$^{3+}$, there are many permutations of ions and hosts in which MD emission can likely be observed.  While further study is needed to find the most practical combinations, these comprehensive calculations provide a solid foundation from which to begin this search, and they provide a firm set of numbers with which to analyze future experimental data. The tabulated values may also be helpful in studying the potential role of MD transitions in more complex processes such as upconversion~\cite{ChanJPhysChemB2012} and quantum cutting~\cite{WeghJLum1999}. These same calculations can also help focus the design of optical structures to enhance MD emission. For example, emission wavelengths, transition rates, and branching ratios can be used as the starting point for simulating the effects of optical antennas and metamaterials on MD transitions. Combining these quantum-mechanical calculations with experimental measurements and electromagnetic simulations can expand the toolkit with which to access the naturally occurring MD transitions of lanthanide ions.

\begin{acknowledgments}
We thank S. Cueff, M. Jiang, S. Karaveli, J. Kurvits and D. Li for fruitful discussions. Financial support was provided by the Air Force Office of Scientific Research (PECASE award FA-9550-10-1-0026) and the National Science Foundation (CAREER award EECS-0846466).
\end{acknowledgments}
\appendix
\section{Free Ion Hamiltonian}
\paragraph*{}Closed form expressions of the interaction terms used in these calculations are provided below. These expressions are well defined through many different publications and are provided here for reference purposes.
\subsection{Coefficients of Fractional Parentage}
\label{CFP App}
\paragraph*{}When describing a particular term in the $l^{\text{n}}$ configuration, one must realize that there could be multiple ways in which to arrive at that term from the $l^{\text{n-1}}$ configuration. There is an approach to this problem that was developed by Giulio Racah,\cite{RacahPhysRev1942I,RacahPhysRev1942II,RacahPhysRev1943,RacahPhysRev1949} which defines the terms of the $l^{\text{n}}$ configuration in terms of $l^{\text{n-1}}$. The terms of $l^{\text{n-1}}$ are known as the parents of the corresponding daughters $l^{\text{n}}$. These coefficients of fractional parentage (CFP) need only be calculated once. For this paper, the CFP were not calculated directly but an electronic version of the tables produced by \citet{NielsonKoster1963} was used instead.\footnote{Available at http://www.pha.jhu.edu/groups/cfp} All subsequent calculations were made using these values. The CFP are denoted by $\left(\psi\{|\psi\right)$. Due to the fact that a particular state might appear in more than one configuration, such as in both the $4f^{\text{n}}$ and $4f^{\text{n+2}}$ configurations, a method to distinguish when a state appears is necessary. This is accomplished by using the seniority number, which can take integer values from $1$ to $7$, indicating in which $4f^{\text{n}}$ configuration a state first appears.
\subsection{Electrostatic Interaction}
\paragraph*{} The electrostatic interaction occurs between configurations with two or more electrons. This is a result of the Coulomb repulsion between the two electrons. It is calculated from two single electron wavefunctions. The electrostatic interaction is diagonal in both J and S values and the matrix elements are found using the following expression:\cite{CowanTheoryAtomic1981}
\begin{eqnarray*}
\left< l^{n}\psi^\prime L^\prime S\left|f_k \right| l^{n}\psi{}LS \right>=\frac{1}{2}\left< l\left|C^{(k)}\right| l\right> ^2 \biglb(\frac{1}{2L+1}\\
\times{}\sum_{\bar{\psi}, \bar{L}}\left<   l^{n}\bar{\psi} \bar{L} S \left| U^{(k)} \right| l^{n}\psi^\prime L^\prime S\right> \left< l^{n}\bar{\psi} \bar{L} S\left|U^{(k)} \right| l^{n}\psi L S \right>\\
-\delta_{\psi \bar{\psi}} \frac{n (4 l +2-n)}{(2 l+1)(4l+1)}\bigrb).
\end{eqnarray*}
$C^{(k)}$ is the irreducible tensor defined by Racah,\cite{RacahPhysRev1942II} and $U^{(k)}$ is the irreducible tensor tabulated by Nielson and Koster.\cite{NielsonKoster1963} Since we are concerned with $f^{\text{n}}$ configurations, we used $l=3$ for all calculations. Again, we are using the notation in which $\psi{}$ represents all other quantum numbers that are not specifically mentioned.

\subsection{Spin-Orbit Interaction}
\paragraph*{} The spin-orbit interaction is, in essence, a dipole-dipole interaction. The spin-orbit interaction is diagonal in $J$ but not in $L$ or $S$. We calculated this interaction using the following formula:
\begin{eqnarray*}
\left< f^{n}\psi^\prime L^\prime S^\prime\left|A_{SO} \right| f^{n}\psi{}LS \right>=(-1)^{J+L+S^\prime} \SixJSymbol{L}{L^\prime}{1}{S^\prime}{S}{J}\nonumber\\
\times \left<f^{n}\psi^\prime L^\prime S^\prime \left| V^{(11)} \right| f^{n}\psi L S\right>.
\end{eqnarray*}
Here we are using the conventional notation for the Racah 6-j symbols and $V^{(11)}$ is the irreducible tensor tabulated by \citet{NielsonKoster1963}. 

\subsection{Two-Body Interaction}
\paragraph*{}For configurations with two or more valence electrons (or holes), $4f^2$ to $4f^{12}$, two-body interactions are used to help correct for the use of single electron wavefunctions. The first term in this correction was discovered by \citet{TreesPhysRev1952}. The other two terms are calculated using the Racah numbers and the Casimir operator $G$.\cite{TreesPhysRev1961} The eigenvalues of the Casimir operator on the groups $R_7$ and $G_2$ can be found in~\citet{WybourneSpecProps1965}. 

\subsection{Three-Body Interaction}
\paragraph*{} The three-body interaction terms are analogous to the two-body but exist for only $4f^3$ to $4f^{11}$. The form of this operator is:\cite{HansenAtomicData1996}
\begin{eqnarray*}
\left<f^{\text{n}} \psi \left| t_i \right| f^{\text{n}} \psi^\prime \right>&=&\frac{n}{n-3}\\
& \times& \sum_{\bar{\psi},\bar{\psi^\prime}} (\psi\{|\bar{\psi}) (\psi^\prime\{|\bar{\psi^\prime}) \left(f^{\text{n-1}}\bar{\psi}\left|t_i\right|f^{\text{n-1}}\bar{\psi^\prime}\right).
\end{eqnarray*}
This operator is built up recursively using the values for the $4f^3$ states found in Tables 1 and 2 of Ref.~\onlinecite{JuddJOSAB1984}.

\subsection{Spin-Spin Interaction}
\paragraph*{} The spin-spin interaction is analogous to the spin-orbit but is the interaction between the spins of two electrons. $H_{ss}$ is calculated recursively, using the reduced matrix operator $T^{(22)}$. $T^{(22)}$ is defined for the $4f^2$ configuration, these defined values then permit the calculation for all $4f^{\text{n}}$, $n\geq{}2$, configurations and using the following equation:\cite{JuddPhysRev1968MagInt}
\begin{eqnarray*}
\left<f^{\text{n}} \psi \left| T^{(22)} \right| f^{\text{n}} \psi^\prime \right>=\delta_{J,J^\prime} (-1)^{S^\prime+L+J}\\
\times \sum_{\bar{\psi},\bar{\psi^\prime}}\left(\psi\{|\psi\right) \SixJSymbol{S^\prime}{L^\prime}{J}{L}{S}{1} \left(f^{\text{n-1}}\bar{\psi}\left|T^{(22)}\right|f^{\text{n-1}}\bar{\psi^\prime}\right).
\end{eqnarray*}

\subsection{Spin-Other-Orbit and Electrostatically Correlated Spin-Orbit Interactions}
\paragraph*{} The spin-other-orbit interaction is an interaction between the spin of one electron and the orbit of another. It is only valid for $4f^2$ to $4f^{12}$ configurations. The electrostatically correlated spin-orbit interaction is a configuration interaction between the spin of an electron in one configuration with the orbit of an electron residing in a different configuration. These terms were grouped together for calculation by Judd, Crosswhite and Crosswhite.\cite{JuddPhysRev1968MagInt} The following form was used:\cite{ChenJofLum2007}
\begin{eqnarray*}
\left<f^{\text{n}} \psi \left| T^{(11)}+t^{(11)}-a z_{13} \right| f^{\text{n}} \psi^\prime \right>=\delta_{J,J^\prime} (-1)^{S^\prime+L+J}\\
\times{}\sum_{\bar{\psi},\bar{\psi^\prime}}\left(\psi\{|\psi\right) \SixJSymbol{S^\prime}{L^\prime}{J}{L}{S}{1}\\
\times{}\left(f^{\text{n-1}}\bar{\psi}\left|T^{(11)}+t^{(11)}-a z_{13}\right|f^{\text{n-1}}\bar{\psi^\prime}\right).
\end{eqnarray*}
Both $T^{(11)}$ and $t^{(11)}$ are reduced matrix operators. These reduced matrix operators in addition to the values $a$ and $z_{13}$ are defined for the $4f^2$ configuration in Refs. \onlinecite{JuddPhysRev1968MagInt} and \onlinecite{ChenJofLum2007}.
\section{Magnetic Dipole Transitions}
\label{MDTransitionsApp}

\subsection{Oscillator Strength}
\paragraph*{} All MD ground state absorption lines were calculated using the following equation:\cite{ShortleyPhysRev1940} 
\[
f_{MD}=\frac{8\pi^2m_e}{3he^2c}\left(\frac{n_r}{\lambda}\right)\frac{1}{2J+1}S_{MD},
\]
where $S_{MD}$ is the magnetic dipole transition line strength. This line strength is defined as:
\[
S_{MD}=\frac{e\hbar}{2m_ec}\sum_{\psi,\psi^\prime}\left|\left<\psi^\prime{}\left|L+g_eS\right|\psi{}\right>\right|^2,
\]
where $g_e$ is the gyromagnetic ratio of the electron. A list of all non-zero absorption lines can be found in the Supplemental Material.\footnotemark[4]
\subsection{Transition Rates}
\paragraph*{} All MD emission lines were calculated using the following equation:\cite{ShortleyPhysRev1940} 
\[
A_{MD}=\frac{1}{2J+1}\frac{16\pi^3\mu_0}{3h}\left(\frac{n_r}{\lambda}\right)^3S_{MD},
\]
and all non-zero transitions can be found in the Supplemental Material.\footnotemark[4]

\vfill
\section{Electric Quadrupole Transitions}
\label{EQTransitionsApp}

\subsection{Oscillator Strength}
\paragraph*{} All EQ ground state absorption lines were calculated using the following equation:\cite{ThorneSpectrophysics1999} \[
f_{EQ}=\frac{112}{225}\frac{\pi^3 a_0^3}{\alpha}\left(\frac{n_r}{\lambda}\right)^3\left< r^2 \right>\frac{S_{EQ}}{2J+1},
\] where $S_{EQ}$ is the electric quadrupole line strength and is defined as:
\begin{eqnarray*}
S_{EQ}=(-1)^{S+L^\prime+J+2}\sqrt{(2J+1)(2J^\prime+1)}\\
\times\SixJSymbol{J}{J^\prime}{2}{L^\prime}{L}{S} \left< \psi^\prime{}\left|U^{(k)} \right| \psi\right>.
\end{eqnarray*}
Calculated values for the expectation value of the radial wavefunctions for the lanthanide series, $\left< r^2 \right>$, were taken from Table 21.8 in Ref.~\onlinecite{WybourneOptSpecLanthidesMagHyper}. A list of all non-zero absorption lines can be found in the Supplemental Material.\footnotemark[4]

\subsection{Transition Rates}
\paragraph*{} All EQ emission lines were calculated using the following equation:\cite{ThorneSpectrophysics1999}
\[
A_{EQ}=\frac{1}{2J+1}\frac{8\pi^5}{5h\epsilon_0}\left(\frac{n_r}{\lambda}\right)^5S_{EQ}.
\]
There are a total of 3079 non-zero EQ transitions between $300$ and $1700$ nm,  all such transitions can be found in the Supplemental Material.\footnotemark[4]
\footnotetext[4]{See Supplemental Material at [URL will be inserted by publisher] for a complete tabulation of all nonzero MD and EQ transitions between 300 and 1700 nm for each trivalent lanthanide ion.}

\end{document}